\documentclass[sigconf]{acmart} 
\usepackage{bm}
\usepackage[linesnumbered,ruled,vlined]{algorithm2e}

\AtBeginDocument{%
  \providecommand\BibTeX{{%
    \normalfont B\kern-0.5em{\scshape i\kern-0.25em b}\kern-0.8em\TeX}}}

\acmYear{2021}\copyrightyear{2021}
\setcopyright{acmlicensed}
\acmConference[e-Energy '21]{The Twelfth ACM International Conference on Future Energy Systems}{June 28--July 2, 2021}{Virtual Event, Italy}
\acmBooktitle{The Twelfth ACM International Conference on Future Energy Systems (e-Energy '21), June 28--July 2, 2021, Virtual Event, Italy}
\acmPrice{15.00}
\acmDOI{10.1145/3447555.3464848}
\acmISBN{978-1-4503-8333-2/21/06}

\begin{document}

\title{Blockchain for Transactive Energy Management of Distributed Energy Resources in Smart Grid}

\author{Qing Yang}
\affiliation{%
  \department{College of Electronics and Information Engineering}
  \institution{Shenzhen University}
  \city{Shenzhen}
  \state{Guangdong}
  \country{China}
  \postcode{518060}
}
\email{yang.qing@szu.edu.cn}

\author{Hao Wang}
\authornote{Corresponding author: \url{hao.wang2@monash.edu} (Hao Wang).}
\affiliation{%
  \department{Department of Data Science and AI}
  \institution{Monash University}
  \city{Melbourne}
  \state{Victoria}
  \country{Australia}
  \postcode{3800}
}
\email{hao.wang2@monash.edu}

\author{Xiaoxiao~Wu, Taotao~Wang, Shengli~Zhang}
\affiliation{%
  \department{College of Electronics and Information Engineering}
  \institution{Shenzhen University}
  \city{Shenzhen}
  \state{Guangdong}
  \country{China}
  \postcode{518060}
}
\email{{xxwu.eesissi,ttwang,zsl }@szu.edu.cn}

\begin{abstract}
This work presents the design and implementation of a blockchain system that enables the trustable transactive energy management for distributed energy resources (DERs). We model the interactions among DERs, including energy trading and flexible appliance scheduling, as a cost minimization problem. Considering the dispersed nature and diverse ownership of DERs, we develop a distributed algorithm to solve the optimization problem using the alternating direction method of multipliers (ADMM) method. Furthermore, we develop a blockchain system, on which we implement the proposed algorithm with the smart contract, to guarantee the transparency and correctness of the energy management. We prototype the blockchain in a small-scale test network and evaluate it through experiments using real-world data. The experimental results validate the feasibility and effectiveness of our design.
\end{abstract}



\maketitle

\section{Introduction}
The fast-growing penetration of distributed energy resources (DERs), such as distributed renewables, energy storage, electric vehicles, and controllable loads, poses significant challenges to the centralized power systems with unidirectional power flow. Successful integration of heterogeneous DERs calls for a paradigm shift to a decentralized power system with bidirectional power flow. 

A large body of literature has focused on the energy management and energy trading of various DERs managed by an aggregator. For example, a service-centric virtual power plant (VPP) was studied in \cite{koraki2017wind} to integrate renewable energy into an electricity market enabled by the cooperation between the VPP and the distribution system operator. A similar work \cite{kasaei2017optimal} proposed to aggregate distributed generators, energy storage, and controllable loads in a VPP to mitigate the impact of the variable generations. More recent research explored various types of services that DERs can provide. For example, as an emerging service, energy trading was explored in \cite{nguyen2018optimizing, alam2019peer}. In \cite{nguyen2018optimizing}, an optimization model was presented to maximize the economic benefits for solar-battery distributed generations in a peer-to-peer (P2P) energy trading scenario. In \cite{alam2019peer}, the authors developed an algorithm to optimize the energy cost for smart homes under P2P energy trading and evaluated the impact of energy trading in a microgrid. However, the above studies adopted a centralized solution to the energy management of DERs, which is not practical as DERs are usually not owned by the aggregator.

Distributed optimization attracted research attention as a promising solution to manage DERs. The alternating direction method of multipliers (ADMM) method has been used to facilitate energy trading among interconnected microgrids \cite{wang2016incentivizing} and smart homes \cite{yang2020cooperative}. A decentralized optimization algorithm was proposed in \cite{li2016admm} to optimize the demand response of electric vehicles. A fully distributed algorithm was proposed in \cite{chen2018fully} using ADMM and the consensus mechanism. The above studies contributed to the distributed or decentralized energy management of DERs. However, the computing and communication process embedded in the distributed algorithm needs a trusted and verifiable computing environment, which becomes one of the major obstacles to implementing distributed transactive energy paradigm.

Blockchain \cite{swan2015blockchain}, as a distributed ledger, can remove the barrier of lacking a trusted and verifiable computing environment. Blockchain is a tamper-proof decentralized ledger maintained by a group of nodes through the consensus algorithm \cite{xiao2020survey}. Furthermore, the blockchain supports the execution of generic computer programs as smart contracts, resulting in the proliferation of various decentralized applications \cite{eth}. Existing studies in \cite{mengelkamp2018designing} deployed a blockchain-based P2P energy trading platform to facilitate online payments for a microgrid in Brooklyn of New York City. In addition, many existing studies employed blockchain technology as a secure and convenient online payment tool \cite{bao2020survey,mollah2020blockchain}.

Our work adopts the blockchain as a trustable computing machine to implement the distributed transactive energy management algorithm for DERs and also as a secure communication and payment tool. We formulate a cost minimization problem to optimize DERs' interactions, including energy trading, energy scheduling, and grid services. We also decompose the cost minimization problem and solve it using ADMM to manage DERs in a distributed fashion. We design a blockchain system as a trustable transactive energy management platform and implement the distributed algorithm in a smart contract, which opens up opportunities for trustable and efficient integration of DERs in the smart grid.

\section{System Model}

\begin{figure}[!t]
    \centering
    \includegraphics[width=8.5cm]{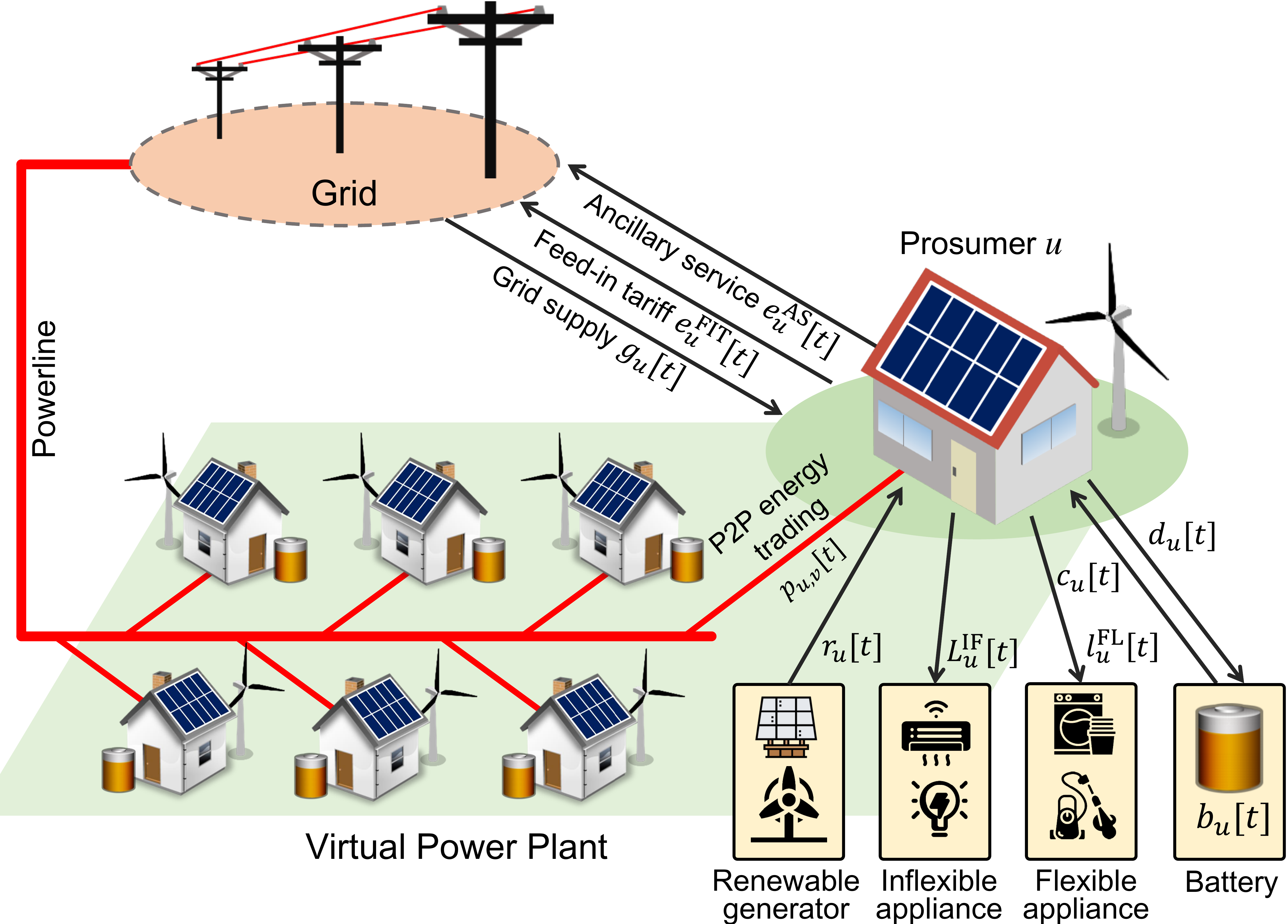}
    \vspace{-0.2cm}
    \caption{The system model of the virtual power plant.}
    \vspace{-0.6cm}
    \label{f1:sysmod}
\end{figure}

\begin{table}[!tb]
    \centering
    \caption{The definitions of the notations.\vspace{-0.3cm}}\label{tab:short}
    \footnotesize
    \renewcommand{\arraystretch}{1.3}
    \label{t1:cost}
    \begin{tabular}{c c l}
        \hline
        \textbf{Variable} & \textbf{Definition} & \textbf{Meaning} \\
        \hline
        $\bm{g}_u$ & $[ g_u[1], \dots, g_u[24] ]^{T}$ & Grid usage of prosumer $u$ \\
        \hline
        $\bm{r}_u$ & $[ r_u[1], \dots, r_u[24] ]^{T}$ & Renewable generation of prosumer $u$ \\
        \hline
        $\bm{l}_u^{\mathrm{FL}}$ & $[ l_u^{\mathrm{FL}}[1], \dots, l_u^{\mathrm{FL}}[24] ]^{T}$ & Flexible appliance of prosumer $u$ \\
        \hline
        $\bm{L}_u^{\mathrm{IF}}$ & $[ L_u^{\mathrm{IF}}[1], \dots, L_u^{\mathrm{IF}}[24] ]^{T}$ & Inflexible appliance of prosumer $u$ \\
        \hline
        $\bm{p}_{u,v}$ & $[ p_{u,v}[1], \dots, p_{u,v}[24] ]^{T}$ & Energy purchased by prosumer $u$ from $v$ \\
        \hline
        $\bm{c}_u$ & $[ c_u[1], \dots, c_u[24] ]^{T}$ & Battery charging of prosumer $u$\\
        \hline
        $\bm{d}_u$ & $[ d_u[1], \dots, d_u[24] ]^{T}$ & Battery discharging of prosumer $u$\\
        \hline
        $\bm{e}_u^{\mathrm{AS}}$ & $[ e_u^{\mathrm{AS}}[1], \dots, e_u^{\mathrm{AS}}[24] ]^{T}$ & prosumer $u$'s ancillary service \\
        \hline
        \vspace{-0.5cm}
    \end{tabular}
\end{table}

We consider a cluster of prosumers owning DERs in a virtual power plant (VPP) system and denote prosumers $u \in \mathcal{U} {=} \{1,\dots,U\}$, where $U$ is the total number of the prosumers as shown in Fig.~\ref{f1:sysmod}. The prosumers have renewable energy generators (e.g., solar panels and small wind turbines) and various in-house appliances such as air conditioners and washers. Each prosumer also installs a battery energy storage system (BESS). The smart meter and the home energy management system manage and schedule the above appliances and energy trading with the outside. We focuses on the day-ahead scheduling and denote the operational horizon by $t \in \mathcal{H} {=} \{1,\dots,24\}$ with one-hour time slot.

We summarize the notations in Tab.~\ref{tab:short}. Note that the boldfaced variables are vectors for one-day profiles. For example, $g_u[t]$ represents the prosumer $u$'s grid power usage in the $t$-th time slot, and $\bm{g}_u = [ g_u[1], \dots, g_u[24] ]^{T}$ represents its grid usage of the whole day. Therefore, prosumer $u$'s electricity charge to the grid operator is $\mathcal{C}_u^{\mathrm{G}} = \alpha \sum_{t\in \mathcal{H}} g_{u}[t] + \beta \max_{t\in \mathcal{H}} g_{u}[t]$,
where $\alpha$ is the energy rate and $\beta$ is the peak electricity rate.

Each prosumer has two types of electric appliances: flexible appliances and inflexible appliances. The flexible appliances, such as the washer and dryer, are time-shiftable, so that they can be scheduled over the operational horizon. By contrast, inflexible appliances (e.g., the house lighting and refrigerator) cannot be adjusted or shifted over time. We assume that the prosumer $u$ has a preferred schedule for the flexible appliances, denoted by $L_u^{\mathrm{Ref}}[t], \forall t \in \mathcal{H}$. Therefore, the deviations from the preferred schedule incurs a discomfort cost for prosumer $u$ denoted by $\mathcal{C}_u^{\mathrm{FL}} {=} \sum_{t {\in} \mathcal{H}} ( l_u^{\mathrm{FL}}[t] {-} L_u^{\mathrm{Ref}}[t] )^{2}, \forall u {\in} \mathcal{U}$.

The future smart grid allows prosumers to freely trade energy with each other to utilize renewable energy better. The system operator sets a fixed price $\pi^{\mathrm{P2P}}$ for the P2P energy trading, which is assumed to be lower than the typical price of the grid. Then prosumer $u$'s payment for P2P energy trading is $\mathcal{C}_u^{\mathrm{P2P}} {=} \sum_{t {\in} \mathcal{H}}  \sum_{v \in \mathcal{U}} \pi^{\mathrm{P2P}} p_{u,v}[t]$. Note that a positive $\mathcal{P}_u^{\mathrm{P2P}}$ indicates prosumer $u$'s overall payment to other prosumers, and a negative $\mathcal{P}_u^{\mathrm{P2P}}$ indicates its overall revenue in the energy trading.

The prosumers' energy storage can store extra energy and discharge to support appliances when needed. The VPP system can provide ancillary service to the grid by aggregating the prosumer's energy storage. Let $b_u[t]$ denote the energy level of prosumer $u$'s battery in time slot $t$. Since $b_u[t]$ is decided by the charge/discharge operations of the energy storage, we have $b_u[t] {=} b_u[t-1] {+} \eta c_u[t] {-} \frac{d_u[t]}{\eta}$, where the coefficient $\eta$ indicates the efficiency of the battery. Because charge/discharge operations degrade the lifespan of the energy storage with a cost of $\mathcal{C}_u^{\mathrm{BA}} {=} \sum_{t\in \mathcal{H}} \left( c_u[t] + d_u[t] \right)$.

The VPP system operator incentivizes the prosumers to reserve energy in their energy storage for load regulation or spinning reserve by paying them with some rewards to provide the ancillary service. The reserved energy can be immediately dispatched from the VPP to the grid for various grid services \citep{melo2016robust}. By storing energy in the battery, prosumer $u$ can receive the reward $\mathcal{R}_u^{\mathrm{AS}} = \sum_{t \in \mathcal{H}} \pi^{\mathrm{AS}}[t] e_u^{\mathrm{AS}}[t]$, where $\pi^{\mathrm{AS}}[t]$ is the reward price in time slot $t$. This price is set by the grid operator based on the forecasted grid load.

\section{Problem Formulation}\label{sec:problem}

We formulate a total cost minimization problem for all prosumers with each prosumer's cost defined as
\begin{equation}
             \mathcal{C}_u = \mathcal{C}_u^{\mathrm{G}} + \mathcal{C}_u^{\mathrm{FL}} + \mathcal{C}_u^{\mathrm{BA}} + \mathcal{C}_u^{\mathrm{P2P}} - \mathcal{R}_u^{\mathrm{AS}}, \label{objective-operatingcost2} 
\end{equation}
which consists of costs for energy supplies, load scheduling, energy storage operation, and ancillary service. 

The energy supply and demand of each prosumer must be balanced, and we have 
\begin{equation}
    \bm{l}_u^{\mathrm{FL}} + \bm{L}_u^{\mathrm{IF}} + \bm{c}_u = \bm{r}_u + \bm{g}_u + \bm{d}_u + \sum\nolimits_{v \in \mathcal{U} \backslash u} \bm{p}_{u,v}, ~ \forall u \in \mathcal{U}. \label{constraint-load12}
\end{equation}

Therefore, the total cost minimization problem of all prosumers in the VPP is formulated as
    \begin{equation}
        \begin{aligned}
            \bm{s}^{*} = & \arg \min_{ \left\{ \bm{g}_u, \bm{r}_u, \bm{l}_u^{\mathrm{FL}}, \bm{c}_u, \bm{d}_u, \bm{p}_{u,v \in \mathcal{U}}, \bm{e}_u^{\mathrm{AS}} \mid_{\forall u \in \mathcal{U}} \right\} } \sum_{\forall u \in \mathcal{U} }\mathcal{C}_u, \label{opt3}\\
            &\mathrm{Subject} \: \mathrm{to} \: \text{the system constraints \eqref{constraint-load12},}
        \end{aligned}
    \end{equation}
where the solution $\bm{s}^{*}$ represents the optimal energy management schedules for all the VPP prosumers. Specifically, $\bm{s}^{*} {=} \{ \bm{s}^{*}_u | u {\in}  \mathcal{U}\}$, where $ \bm{s}^{*}_u {\triangleq} (\breve{\bm{g}}_u, \breve{\bm{r}}_u,  \breve{\bm{l}}_u^{\mathrm{FL}}, \breve{\bm{c}}_u, \breve{\bm{d}}_u, \breve{\bm{e}}_u^{\mathrm{AS}})$ represents prosumer $u$'s optimal energy schedule on grid usage, renewable generation, flexible appliance scheduling, energy storage operation, and ancillary service, respectively. 

The conventional VPP system employs a central coordinator who collects all the prosumers' information and solves the optimization problem of \eqref{opt3} in a centralized manner. This method, however, has three major drawbacks that impediment its feasibility. First, the effectiveness of this method depends on the correctness of the coordinator, which incurs the risk of single-point failure. Second, it incurs significant privacy concerns because the prosumers have to reveal all of their private information regarding DERs to the coordinator. Third, the solution cannot be verified and trusted by the prosumers because the operation of the coordinator is a ``black box'' to them.

To address these drawbacks, we design a distributed transactive energy management algorithm that can attain the optimal solution while preserving prosumers' privacy. Furthermore, by implementing the proposed algorithm in smart contract on the blockchain, we guarantee the correctness of the results and remove the need for a centralized coordinator.

We employ the primal-dual-type method \citep{jakovetic2020primal} to solve the optimization problem in \eqref{opt3}. Based on ADMM \cite{boyd2011distributed}, we introduce auxiliary variables ${p}^{\prime}_{u,v}[t]$ for prosumer $u$'s energy trading decisions in time slot $t$, and dual variables $\lambda_{u,v}$. The original optimization problem in \eqref{opt3} can be written as two subproblems below.

\begin{figure}[!t]
    \centering
    \includegraphics[width=8cm]{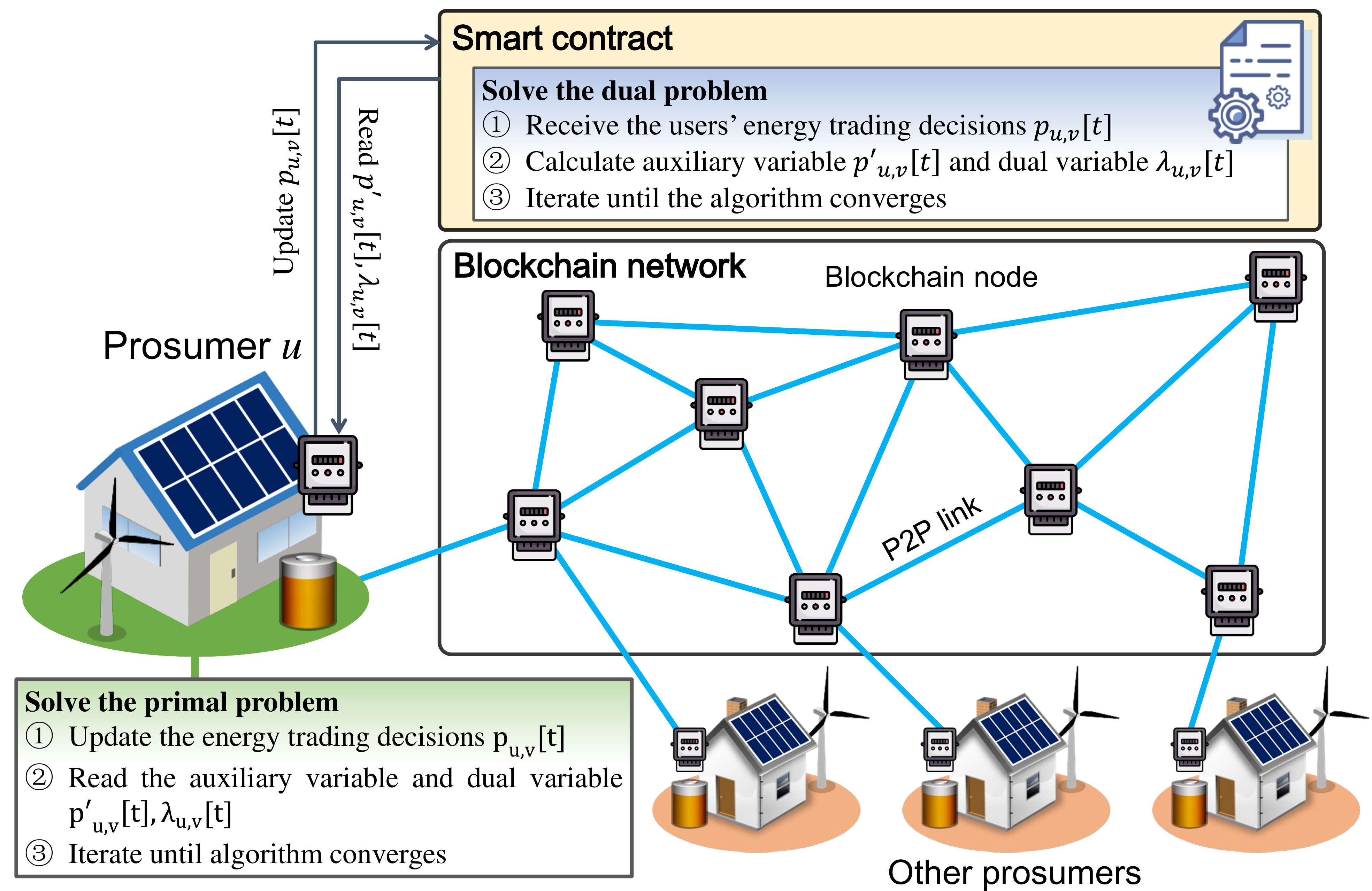}
    \vspace{-0.3cm}
    \caption{The implementation of the decentralized algorithm on blockchain.\vspace{-0.5cm}}
    \vspace{-0.6cm}
    \label{f:process}
\end{figure}

\noindent\textbf{The primal problem}:
\begin{equation}
\begin{aligned}
    &\min \Bigg\{ \mathcal{C}_u {+} \sum_{v \in \mathcal{U}} \sum_{t\in\mathcal{H}} \Big[ \frac{\rho}{2} \left( p^{\prime}_{u,v}[t] {-} p_{u,v}[t] \right)^{2} {-} \lambda_{u,v}[t] p_{u,v}[t] \Big] \Bigg\}, \\
    &\mathrm{with} \: \mathrm{variables:} \: \bm{g}_u, \bm{r}_u, \bm{l}_u^{\mathrm{FL}}, \bm{c}_u, \bm{d}_u, \bm{p}_{u,v \in \mathcal{U}}, \bm{e}_u^{\mathrm{AS}}. \label{primal}
\end{aligned}
\end{equation}
\noindent\textbf{The dual problem}:
\begin{equation}
\begin{aligned}
    &\min \sum_{u\in\mathcal{U}} \sum_{v \in \mathcal{U}} \sum_{t\in\mathcal{H}} \Big[ \frac{\rho}{2} \left( {p}^{\prime}_{u,v}[t] - {p}_{u,v}[t] \right)^{2} + \lambda_{u,v}[t] {p}^{\prime}_{u,v}[t]  \Big]\\
    & \mathrm{with} \: \mathrm{variables:} \: {p}^{\prime}_{u,v}[t], \forall u,v \in \mathcal{U}, \forall t \in \mathcal{H}, \label{dual}
\end{aligned}
\end{equation}
where the coefficient ${\rho}$ denotes the penalty sensitivity of the auxiliary variables.

The primal-dual method works in an iterative manner. The primal problem updates the prosumers' decisions, including the energy trading decisions ${p}_{u,v}[t]$ and other scheduling decisions. The dual problem uses ${p}_{u,v}[t]$ to update the auxiliary variables ${p}^{\prime}_{u,v}[t]$ and the dual variables $\lambda_{u,v}[t]$, which are used by the primal problem in the next iteration. We measure the convergence error by the Euclidean distance between the auxiliary variable $\bm{p}^{\prime}_{u,v}$ and its original variable $\bm{p}_{u,v}$. Since the optimization problem in \eqref{opt3} is convex, the algorithm solving primal and dual problems can converge to the optimal solution.

\section{System Implementation}
\subsection{Design of the Blockchain System}
We build the underlying blockchain system based on the source code of Ethereum Release 1.7 \cite{geth}. Ethereum is a well-verified mainstream blockchain project that has been applied in many areas. It is open-source so that we can modify its source code to tailor the blockchain to smart grid scenarios. Furthermore, Ethereum enables us to implement the energy management algorithm in the previous section as the smart contract. 

The prosumer's smart meter runs the blockchain software to become a blockchain node, since the smart meter is an embedded device that supports the operating system and application software. The smart meters connect to the blockchain network via the communication links such as LoRa and 5G Narrowband IoT. As shown in Fig.~\ref{f:process}, the blockchain nodes form a peer-to-peer network that supports data communication and the operation of the blockchain.

The consensus protocol, which is used to synchronized the states of all the blockchain nodes, has a critical impact on the performance of the blockchain. To run the blockchain on the smart meter, we modified the consensus protocol from PoW (proof of work) to PoA (proof of authority) because the computational complexity of PoW is prohibitively high for the smart meters hardware. The blockchain nodes in Fig.~\ref{f:process} can be divided into \emph{validators} and \emph{normal nodes}. The validators form a committee to receive the transactions, execute smart contracts, and package them to generate a new block in a round-robin manner. The validators can also vote to add a normal node into or remove a validator out of the committee. The normal nodes are the normal prosumers that can send transactions and interact with the smart contract. The normal nodes can access the data on the blockchain and verify the execution of the transactions.

\subsection{Decentralized Algorithm}

As shown in Fig.~\ref{f:process}, the implementation of the decentralized algorithm aims to solve the primal-dual problem in Section~3. The primal problem is locally solved by the prosumers' smart meters using the quadratic programming package of the GNU Octave \citep{octave}; therefore, the prosumers' private energy schedule information is protected. The dual problem is solved by the smart contract that is deployed by the VPP operator on the blockchain, which guarantees the transparency and correctness of the results. The prosumers can exchange the value of auxiliary variables with the smart contract by sending transactions to call the smart contract functions.

We implement the smart contract in Solidity \citep{solidity} to provide three core functions. \texttt{Func A} solves the dual optimization problem. Note here we implement \texttt{Func A} as a pre-defined function in Go language because Solidity does not support floating-point computation yet. \texttt{Func B} sets new values to the variables that store the prosumers' energy trading decisions $\bm{p}_{u,v}$. The prosumers can call this function to update their local trading decisions in each iteration. \texttt{Func C} outputs the values of the dual variables $\bm{\lambda}_{u,v}$ and the auxiliary variables $\bm{p}^{\prime}_{u,v}$ computed by the first function. The prosumers can call this function to read the latest values of $\bm{\lambda}_{u,v}$ and  $\bm{p}^{\prime}_{u,v}$ in each iteration. The complete process of the algorithm is listed in Algorithm~\ref{a:1} in the appendix.

\section{Evaluation}
\begin{figure}[!t]
    \centering
    \includegraphics[width=8.5cm]{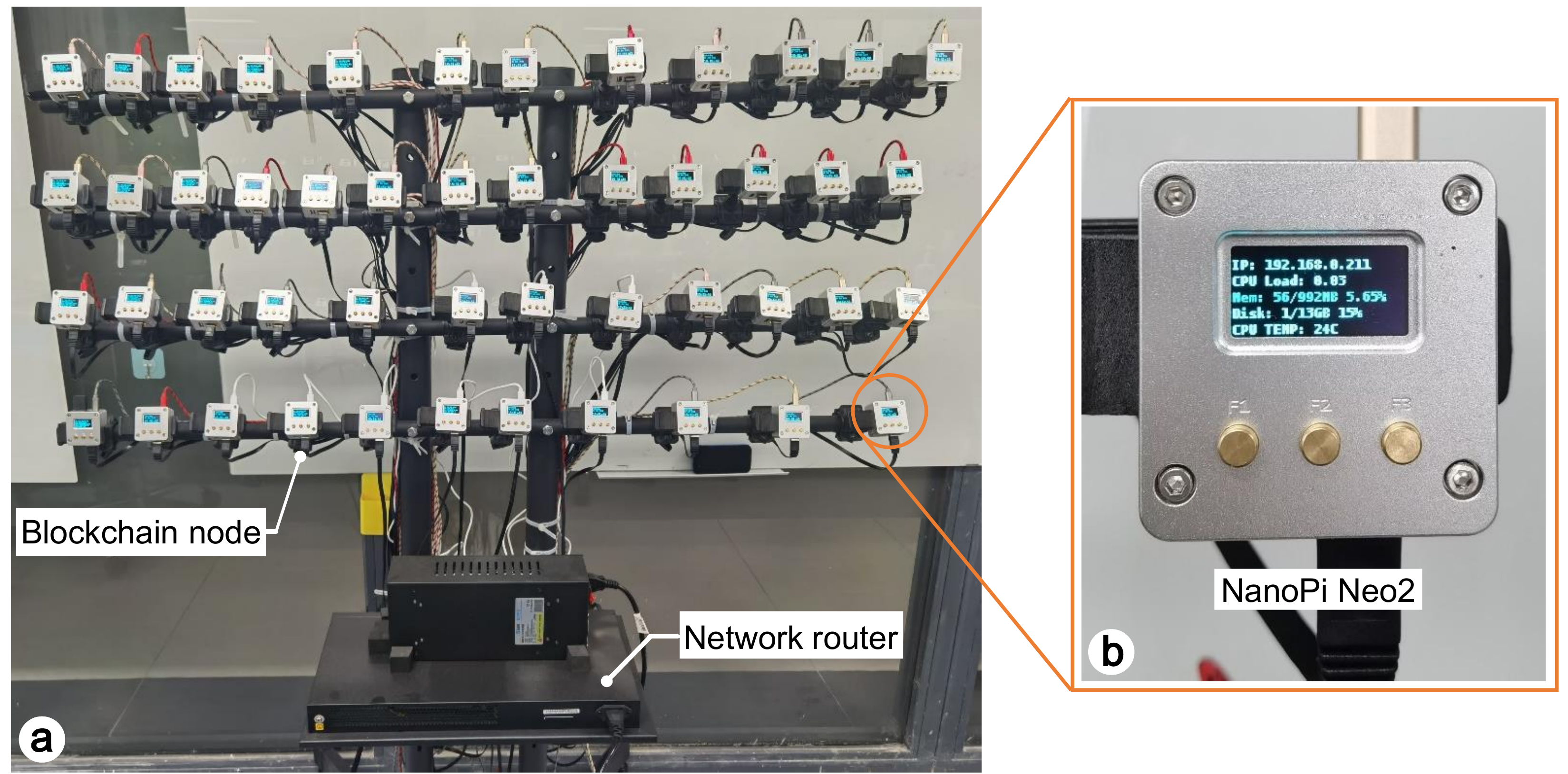}
    \vspace{-0.3cm}
    \caption{System implementation: (a) The test network; (b) The NanoPi Neo2 hardware.}
    \vspace{-0.4cm}
    \label{f:hw}
\end{figure}

\begin{figure}[!t]
    \centering
    \includegraphics[width=8.5cm]{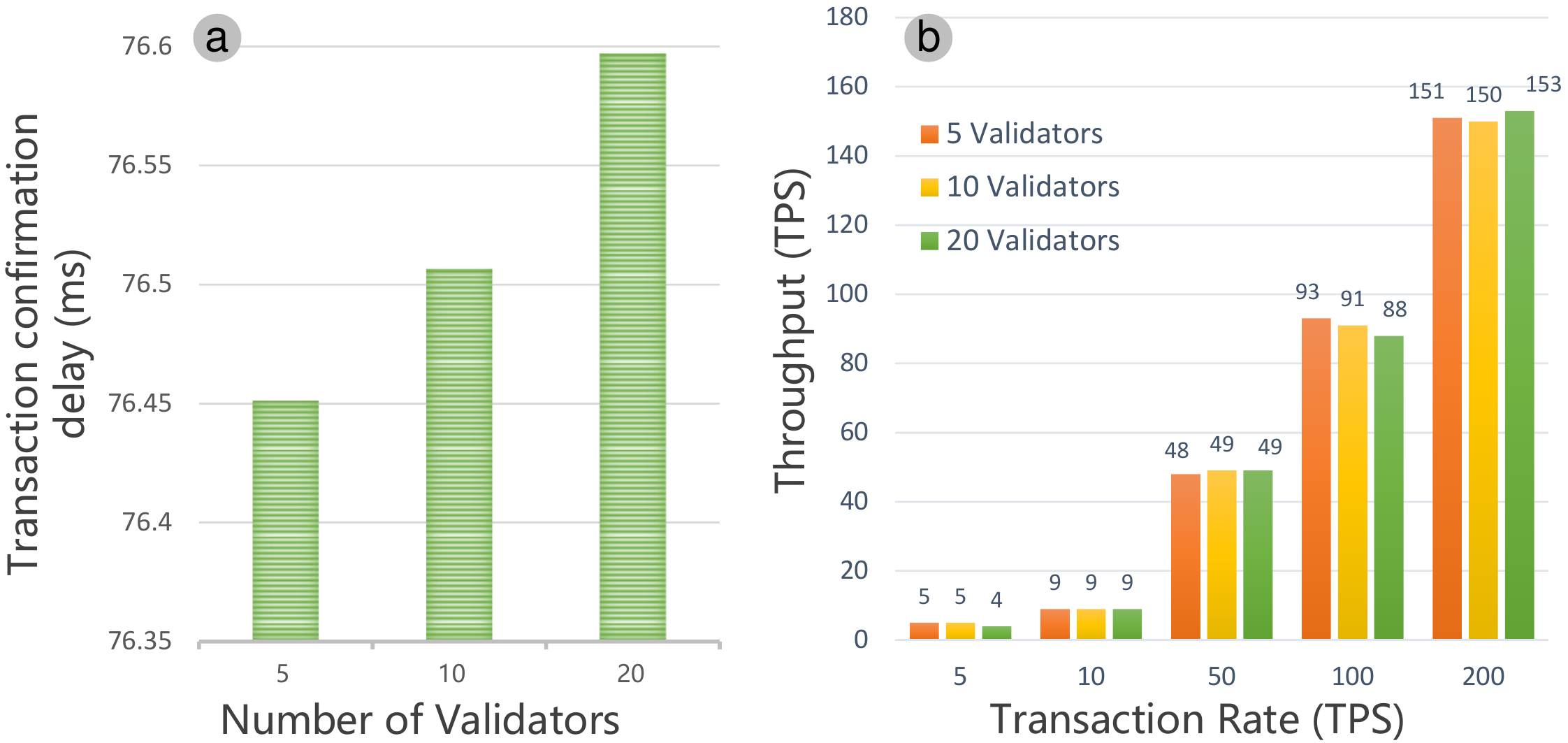}
    \vspace{-0.3cm}
    \caption{Blockchain performance: (a) Transaction confirmation delay; (b) Throughput.}
    \vspace{-0.2cm}
    \label{f:tps}
\end{figure}

To evaluate the performance of the blockchain system, we build a test network with 48 NanoPi Neo2 \citep{nanopi} as shown in Fig.~\ref{f:hw}. The details of the test network are described in Appendix~A.1. We measure the transaction confirmation delay (TCD) and throughput of the blockchain system with different numbers of validators (5, 10, and 20). The TCD measures the time that elapses from the moment when the node transmits a transaction to the moment when the block containing this transaction is decided by the consensus algorithm. Fig.~\ref{f:tps}a shows that the TCDs with different numbers of validators are quite close and all below $76.6$ms. The throughput measured in transactions per second (TPS) is shown in Fig.~\ref{f:tps}b. In the experiment, we gradually increase the transaction transmit rate and measure the average number of confirmed transactions. We observe that the throughput increases and saturated at $150$TPS, which is the peak throughput of the blockchain. The experimental result shows that the underlying blockchain is able to support the execution of the decentralized transactive energy management algorithm.

\begin{figure}[!t]
    \centering
    \includegraphics[width=8.5cm]{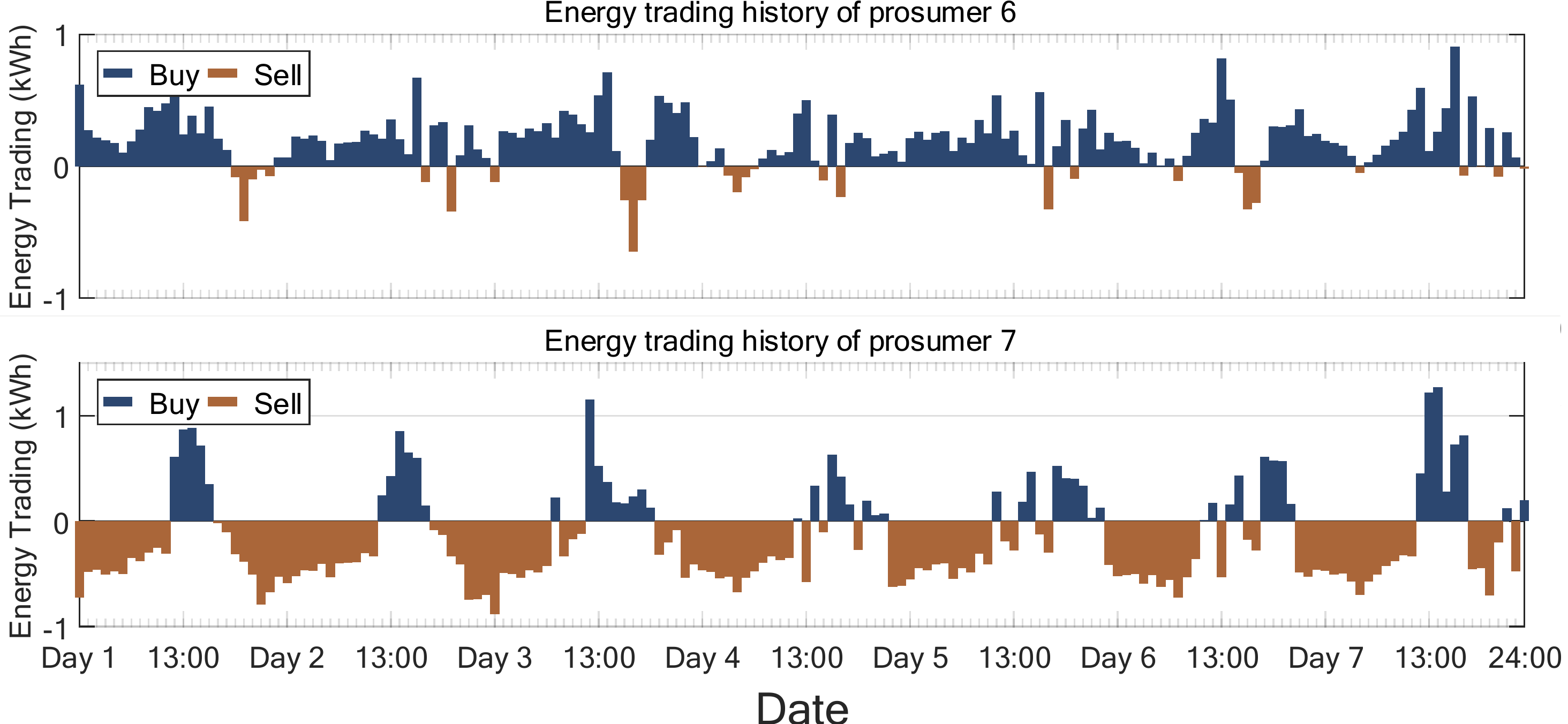}
    \vspace{-0.5cm}
    \caption{The energy trading records of two typical prosumers: prosumer 6 and prosumer 7.}
    \vspace{-0.3cm}
    \label{f:sch}
\end{figure}

To evaluate the performance of the Algorithm~\ref{a:1}, we collect the energy usage data from the real-world smart grid system, including solar/wind power generations \citep{wang2015joint} and in-house energy consumption records \citep{pecan}. We let 10 NanoPis be the prosumers and simulate for seven days. Fig.~\ref{f:sch} shows the optimal energy trading of two typical prosumers. Prosumer 7 has higher renewable generations and sells a lot to others, whereas prosumer 6 is short of local renewables hence frequently buys energy from other prosumers.

\section{Conclusion}
We developed a blockchain system for the transactive energy management of distributed energy resources in a smart grid. We considered a comprehensive model of DERs and employed the ADMM to design a decentralized optimization algorithm. To guarantee the transparency and correctness of the algorithm, we designed a blockchain system tailored for smart meters, and implemented the decentralized algorithm in smart contract. Moreover, we build a test network of 48 NanoPis to validate the blockchain and use real-world data to evaluate the blockchain-based transactive energy management algorithm. Experiments showed that the blockchain can achieve a throughput of 150TPS with a latency below 76ms, and the transactive energy management algorithm could improve the efficiency of the smart grid.

\begin{acks}
This work is in part supported by the National Natural Science Foundation of China (project no.61901280) and the FIT Academic Staff Funding of Monash University.
\end{acks}

\newpage
\bibliographystyle{ACM-Reference-Format}
\bibliography{ref.bib}


\begin{thebibliography}{23}


\ifx \showCODEN    \undefined \def \showCODEN     #1{\unskip}     \fi
\ifx \showDOI      \undefined \def \showDOI       #1{#1}\fi
\ifx \showISBNx    \undefined \def \showISBNx     #1{\unskip}     \fi
\ifx \showISBNxiii \undefined \def \showISBNxiii  #1{\unskip}     \fi
\ifx \showISSN     \undefined \def \showISSN      #1{\unskip}     \fi
\ifx \showLCCN     \undefined \def \showLCCN      #1{\unskip}     \fi
\ifx \shownote     \undefined \def \shownote      #1{#1}          \fi
\ifx \showarticletitle \undefined \def \showarticletitle #1{#1}   \fi
\ifx \showURL      \undefined \def \showURL       {\relax}        \fi
\providecommand\bibfield[2]{#2}
\providecommand\bibinfo[2]{#2}
\providecommand\natexlab[1]{#1}
\providecommand\showeprint[2][]{arXiv:#2}

\bibitem[\protect\citeauthoryear{Alam, St-Hilaire, and Kunz}{Alam
  et~al\mbox{.}}{2019}]%
        {alam2019peer}
\bibfield{author}{\bibinfo{person}{Muhammad~Raisul Alam}, \bibinfo{person}{Marc
  St-Hilaire}, {and} \bibinfo{person}{Thomas Kunz}.}
  \bibinfo{year}{2019}\natexlab{}.
\newblock \showarticletitle{Peer-to-peer energy trading among smart homes}.
\newblock \bibinfo{journal}{\emph{Applied energy}}  \bibinfo{volume}{238}
  (\bibinfo{year}{2019}), \bibinfo{pages}{1434--1443}.
\newblock


\bibitem[\protect\citeauthoryear{Bao, He, Luo, and Choo}{Bao
  et~al\mbox{.}}{2020}]%
        {bao2020survey}
\bibfield{author}{\bibinfo{person}{Jiabin Bao}, \bibinfo{person}{Debiao He},
  \bibinfo{person}{Min Luo}, {and} \bibinfo{person}{Kim-Kwang~Raymond Choo}.}
  \bibinfo{year}{2020}\natexlab{}.
\newblock \showarticletitle{A Survey of Blockchain Applications in the Energy
  Sector}.
\newblock \bibinfo{journal}{\emph{IEEE Systems Journal}}
  (\bibinfo{year}{2020}), \bibinfo{pages}{1--12}.
\newblock
\urldef\tempurl%
\url{https://doi.org/10.1109/JSYST.2020.2998791}
\showDOI{\tempurl}


\bibitem[\protect\citeauthoryear{Boyd, Parikh, Chu, Peleato, and Eckstein}{Boyd
  et~al\mbox{.}}{2011}]%
        {boyd2011distributed}
\bibfield{author}{\bibinfo{person}{Stephen Boyd}, \bibinfo{person}{Neal
  Parikh}, \bibinfo{person}{Eric Chu}, \bibinfo{person}{Borja Peleato}, {and}
  \bibinfo{person}{Jonathan Eckstein}.} \bibinfo{year}{2011}\natexlab{}.
\newblock \showarticletitle{Distributed optimization and statistical learning
  via the alternating direction method of multipliers}.
\newblock \bibinfo{journal}{\emph{Found. Trends Mach. Learn.}}
  \bibinfo{volume}{3}, \bibinfo{number}{1} (\bibinfo{year}{2011}),
  \bibinfo{pages}{1--122}.
\newblock


\bibitem[\protect\citeauthoryear{Chen and Li}{Chen and Li}{2018}]%
        {chen2018fully}
\bibfield{author}{\bibinfo{person}{Guo Chen} {and} \bibinfo{person}{Jueyou
  Li}.} \bibinfo{year}{2018}\natexlab{}.
\newblock \showarticletitle{A fully distributed ADMM-based dispatch approach
  for virtual power plant problems}.
\newblock \bibinfo{journal}{\emph{Applied Mathematical Modelling}}
  \bibinfo{volume}{58} (\bibinfo{year}{2018}), \bibinfo{pages}{300--312}.
\newblock


\bibitem[\protect\citeauthoryear{Elec}{Elec}{2018}]%
        {nanopi}
\bibfield{author}{\bibinfo{person}{Friendly Elec}.}
  \bibinfo{year}{2018}\natexlab{}.
\newblock \bibinfo{booktitle}{\emph{NanoPi Neo2}}.
\newblock
\urldef\tempurl%
\url{https://www.friendlyarm.com/index.php?route=product/product&path=69&product_id=180}
\showURL{%
Retrieved Oct. 1, 2020 from \tempurl}


\bibitem[\protect\citeauthoryear{Foundation}{Foundation}{2017a}]%
        {geth}
\bibfield{author}{\bibinfo{person}{Ethereum Foundation}.}
  \bibinfo{year}{2017}\natexlab{a}.
\newblock \bibinfo{booktitle}{\emph{Go {E}thereum: Official Golang
  implementation of the Ethereum protocol}}.
\newblock
\urldef\tempurl%
\url{https://github.com/ethereum/go-ethereum}
\showURL{%
Retrieved Oct. 30, 2020 from \tempurl}
\newblock
\shownote{Release/1.7.}


\bibitem[\protect\citeauthoryear{Foundation}{Foundation}{2017b}]%
        {solidity}
\bibfield{author}{\bibinfo{person}{Ethereum Foundation}.}
  \bibinfo{year}{2017}\natexlab{b}.
\newblock \bibinfo{booktitle}{\emph{The Solidity Language Documentations}}.
\newblock
\urldef\tempurl%
\url{https://solidity.readthedocs.io/en/v0.5.12/}
\showURL{%
Retrieved Dec. 1, 2019 from \tempurl}
\newblock
\shownote{Release v0.5.12.}


\bibitem[\protect\citeauthoryear{Jakoveti{\'c}, Bajovi{\'c}, Xavier, and
  Moura}{Jakoveti{\'c} et~al\mbox{.}}{2020}]%
        {jakovetic2020primal}
\bibfield{author}{\bibinfo{person}{Du{\v{s}}an Jakoveti{\'c}},
  \bibinfo{person}{Dragana Bajovi{\'c}}, \bibinfo{person}{Jo{\~a}o Xavier},
  {and} \bibinfo{person}{Jos{\'e}~MF Moura}.} \bibinfo{year}{2020}\natexlab{}.
\newblock \showarticletitle{Primal--Dual Methods for Large-Scale and
  Distributed Convex Optimization and Data Analytics}.
\newblock \bibinfo{journal}{\emph{Proc. IEEE}} \bibinfo{volume}{108},
  \bibinfo{number}{11} (\bibinfo{year}{2020}), \bibinfo{pages}{1923--1938}.
\newblock


\bibitem[\protect\citeauthoryear{Kasaei, Gandomkar, and Nikoukar}{Kasaei
  et~al\mbox{.}}{2017}]%
        {kasaei2017optimal}
\bibfield{author}{\bibinfo{person}{Mohammad~Javad Kasaei},
  \bibinfo{person}{Majid Gandomkar}, {and} \bibinfo{person}{Javad Nikoukar}.}
  \bibinfo{year}{2017}\natexlab{}.
\newblock \showarticletitle{Optimal management of renewable energy sources by
  virtual power plant}.
\newblock \bibinfo{journal}{\emph{Renewable energy}}  \bibinfo{volume}{114}
  (\bibinfo{year}{2017}), \bibinfo{pages}{1180--1188}.
\newblock


\bibitem[\protect\citeauthoryear{Koraki and Strunz}{Koraki and Strunz}{2017}]%
        {koraki2017wind}
\bibfield{author}{\bibinfo{person}{Despina Koraki} {and} \bibinfo{person}{Kai
  Strunz}.} \bibinfo{year}{2017}\natexlab{}.
\newblock \showarticletitle{Wind and solar power integration in electricity
  markets and distribution networks through service-centric virtual power
  plants}.
\newblock \bibinfo{journal}{\emph{IEEE Transactions on Power Systems}}
  \bibinfo{volume}{33}, \bibinfo{number}{1} (\bibinfo{year}{2017}),
  \bibinfo{pages}{473--485}.
\newblock


\bibitem[\protect\citeauthoryear{Li, Guo, Sun, and Su}{Li
  et~al\mbox{.}}{2016}]%
        {li2016admm}
\bibfield{author}{\bibinfo{person}{Zhengshuo Li}, \bibinfo{person}{Qinglai
  Guo}, \bibinfo{person}{Hongbin Sun}, {and} \bibinfo{person}{Hangli Su}.}
  \bibinfo{year}{2016}\natexlab{}.
\newblock \showarticletitle{ADMM-based decentralized demand response method in
  electric vehicle virtual power plant}. In \bibinfo{booktitle}{\emph{2016 IEEE
  Power and Energy Society General Meeting (PESGM)}}.
  \bibinfo{publisher}{IEEE}, \bibinfo{address}{Boston, MA, USA},
  \bibinfo{pages}{1--5}.
\newblock


\bibitem[\protect\citeauthoryear{Melo, Trippe, Gooi, and Massier}{Melo
  et~al\mbox{.}}{2016}]%
        {melo2016robust}
\bibfield{author}{\bibinfo{person}{Dante F~Recalde Melo},
  \bibinfo{person}{Annette Trippe}, \bibinfo{person}{Hoay~Beng Gooi}, {and}
  \bibinfo{person}{Tobias Massier}.} \bibinfo{year}{2016}\natexlab{}.
\newblock \showarticletitle{Robust electric vehicle aggregation for ancillary
  service provision considering battery aging}.
\newblock \bibinfo{journal}{\emph{IEEE Transactions on Smart Grid}}
  \bibinfo{volume}{9}, \bibinfo{number}{3} (\bibinfo{year}{2016}),
  \bibinfo{pages}{1728--1738}.
\newblock


\bibitem[\protect\citeauthoryear{Mengelkamp, G{\"a}rttner, Rock, Kessler,
  Orsini, and Weinhardt}{Mengelkamp et~al\mbox{.}}{2018}]%
        {mengelkamp2018designing}
\bibfield{author}{\bibinfo{person}{Esther Mengelkamp},
  \bibinfo{person}{Johannes G{\"a}rttner}, \bibinfo{person}{Kerstin Rock},
  \bibinfo{person}{Scott Kessler}, \bibinfo{person}{Lawrence Orsini}, {and}
  \bibinfo{person}{Christof Weinhardt}.} \bibinfo{year}{2018}\natexlab{}.
\newblock \showarticletitle{Designing microgrid energy markets: A case study:
  The {Brooklyn} Microgrid}.
\newblock \bibinfo{journal}{\emph{Elsevier Appl. Energy}}
  \bibinfo{volume}{210} (\bibinfo{year}{2018}), \bibinfo{pages}{870--880}.
\newblock


\bibitem[\protect\citeauthoryear{Mollah, Zhao, Niyato, Lam, Zhang, Ghias, Koh,
  and Yang}{Mollah et~al\mbox{.}}{2020}]%
        {mollah2020blockchain}
\bibfield{author}{\bibinfo{person}{Muhammad~Baqer Mollah}, \bibinfo{person}{Jun
  Zhao}, \bibinfo{person}{Dusit Niyato}, \bibinfo{person}{Kwok-Yan Lam},
  \bibinfo{person}{Xin Zhang}, \bibinfo{person}{Amer~MYM Ghias},
  \bibinfo{person}{Leong~Hai Koh}, {and} \bibinfo{person}{Lei Yang}.}
  \bibinfo{year}{2020}\natexlab{}.
\newblock \showarticletitle{Blockchain for future smart grid: A comprehensive
  survey}.
\newblock \bibinfo{journal}{\emph{IEEE Internet of Things Journal}}
  \bibinfo{volume}{8}, \bibinfo{number}{1} (\bibinfo{year}{2020}),
  \bibinfo{pages}{18--43}.
\newblock


\bibitem[\protect\citeauthoryear{Nguyen, Peng, Sokolowski, Alahakoon, and
  Yu}{Nguyen et~al\mbox{.}}{2018}]%
        {nguyen2018optimizing}
\bibfield{author}{\bibinfo{person}{Su Nguyen}, \bibinfo{person}{Wei Peng},
  \bibinfo{person}{Peter Sokolowski}, \bibinfo{person}{Damminda Alahakoon},
  {and} \bibinfo{person}{Xinghuo Yu}.} \bibinfo{year}{2018}\natexlab{}.
\newblock \showarticletitle{Optimizing rooftop photovoltaic distributed
  generation with battery storage for peer-to-peer energy trading}.
\newblock \bibinfo{journal}{\emph{Applied Energy}}  \bibinfo{volume}{228}
  (\bibinfo{year}{2018}), \bibinfo{pages}{2567--2580}.
\newblock


\bibitem[\protect\citeauthoryear{Octave}{Octave}{1993}]%
        {octave}
\bibfield{author}{\bibinfo{person}{Octave}.} \bibinfo{year}{1993}\natexlab{}.
\newblock \bibinfo{booktitle}{\emph{GNU Octave: Scientific Programming
  Language}}.
\newblock
\urldef\tempurl%
\url{https://www.gnu.org/software/octave/}
\showURL{%
Retrieved Oct. 1, 2020 from \tempurl}
\newblock
\shownote{Release v4.0.1.}


\bibitem[\protect\citeauthoryear{{Pecan Street}}{{Pecan Street}}{2019}]%
        {pecan}
\bibfield{author}{\bibinfo{person}{{Pecan Street}}.}
  \bibinfo{year}{2019}\natexlab{}.
\newblock \bibinfo{booktitle}{\emph{Pecan Street Dataport}}.
\newblock Pecan Street Inc.
\newblock
\urldef\tempurl%
\url{https://www.pecanstreet.org/dataport/}
\showURL{%
\tempurl}


\bibitem[\protect\citeauthoryear{Swan}{Swan}{2015}]%
        {swan2015blockchain}
\bibfield{author}{\bibinfo{person}{Melanie Swan}.}
  \bibinfo{year}{2015}\natexlab{}.
\newblock \bibinfo{booktitle}{\emph{Blockchain: Blueprint for A New Economy}}.
\newblock \bibinfo{publisher}{O'Reilly Media Inc.},
  \bibinfo{address}{Sebastopol, CA}.
\newblock


\bibitem[\protect\citeauthoryear{Wang and Huang}{Wang and Huang}{2015}]%
        {wang2015joint}
\bibfield{author}{\bibinfo{person}{Hao Wang} {and} \bibinfo{person}{Jianwei
  Huang}.} \bibinfo{year}{2015}\natexlab{}.
\newblock \showarticletitle{Joint investment and operation of microgrid}.
\newblock \bibinfo{journal}{\emph{IEEE Transactions on Smart Grid}}
  \bibinfo{volume}{8}, \bibinfo{number}{2} (\bibinfo{year}{2015}),
  \bibinfo{pages}{833--845}.
\newblock


\bibitem[\protect\citeauthoryear{Wang and Huang}{Wang and Huang}{2016}]%
        {wang2016incentivizing}
\bibfield{author}{\bibinfo{person}{Hao Wang} {and} \bibinfo{person}{Jianwei
  Huang}.} \bibinfo{year}{2016}\natexlab{}.
\newblock \showarticletitle{Incentivizing energy trading for interconnected
  microgrids}.
\newblock \bibinfo{journal}{\emph{IEEE Transactions on Smart Grid}}
  \bibinfo{volume}{9}, \bibinfo{number}{4} (\bibinfo{year}{2016}),
  \bibinfo{pages}{2647--2657}.
\newblock


\bibitem[\protect\citeauthoryear{Wood}{Wood}{2019}]%
        {eth}
\bibfield{author}{\bibinfo{person}{Gavin Wood}.}
  \bibinfo{year}{2019}\natexlab{}.
\newblock \bibinfo{title}{Ethereum: A Secure Decentralized Generalized
  Transaction Ledger}.
\newblock \bibinfo{howpublished}{Yellow Paper}.
\newblock
\urldef\tempurl%
\url{https://ethereum.github.io/yellowpaper/paper.pdf}
\showURL{%
\tempurl}


\bibitem[\protect\citeauthoryear{Xiao, Zhang, Lou, and Hou}{Xiao
  et~al\mbox{.}}{2020}]%
        {xiao2020survey}
\bibfield{author}{\bibinfo{person}{Yang Xiao}, \bibinfo{person}{Ning Zhang},
  \bibinfo{person}{Wenjing Lou}, {and} \bibinfo{person}{Y~Thomas Hou}.}
  \bibinfo{year}{2020}\natexlab{}.
\newblock \showarticletitle{A survey of distributed consensus protocols for
  blockchain networks}.
\newblock \bibinfo{journal}{\emph{IEEE Communications Surveys \& Tutorials}}
  \bibinfo{volume}{22}, \bibinfo{number}{2} (\bibinfo{year}{2020}),
  \bibinfo{pages}{1432--1465}.
\newblock


\bibitem[\protect\citeauthoryear{Yang and Wang}{Yang and Wang}{2020}]%
        {yang2020cooperative}
\bibfield{author}{\bibinfo{person}{Qing Yang} {and} \bibinfo{person}{Hao
  Wang}.} \bibinfo{year}{2020}\natexlab{}.
\newblock \showarticletitle{Cooperative energy management of hvac via
  transactive energy}. In \bibinfo{booktitle}{\emph{2020 IEEE 16th
  International Conference on Control \& Automation (ICCA)}}.
  \bibinfo{publisher}{IEEE}, \bibinfo{address}{Sapporo, Hokkaido, Japan},
  \bibinfo{pages}{1271--1277}.
\newblock


\end{thebibliography}

\appendix
\section{Implementation Details}
\subsection{Hardware Configuration of the Test Network}
We build a small-scale test network in Fig.~\ref{f:hw}a to evaluate the feasibility and performance of the proposed blockchain-based transactive energy management method. We use the NanoPi Neo2 in Fig.~\ref{f:hw}b to emulate the hardware of the smart meter. The NanoPi Neo2 is a low-power industrial embedded device with ARM-A53 (quad-core 1.5GHz) CPU, 1GB memory, and 16GB SD card, which is similar to the hardware configuration of a smart meter. We connect 48 NanoPis by a router to set up the test network. Each NanoPi runs a Ubuntu Core 16.04 operating system and the modified Ethereum client software (based on Release~1.7).

To run the blockchain on the NanoPi Neo2, we modified the consensus protocol from PoW (proof of work) to PoA (proof of authority) to reduce its computational complexity. In our test, running a PoW node on the NanoPi consumes 100\% of the quad-core CPU time and all of its memory, which heavily slows down the operating system. By contrast, a PoA validator node consumes about 1.2\% CPU time, 565MB memory, and 157MB storage (when block height is 33846); a PoA normal node consumes about 0.7\% CPU time, 390MB memory, and 23MB storage. Therefore, the hardware requirements can be satisfied by inexpensive embedded devices and modern smart meters.

\subsection{Pseudocode of the Decentralized Algorithm}
\begin{algorithm}[!h]
     \caption{Transactive energy management} \label{a:1} 
     \SetAlgoLined

     \KwIn{
      dual variable $\bm{\lambda}^{(0)}_{u,v}$, auxiliary variable $\bm{p}^{\prime}_{u,v}$.}
     \KwOut{
     optimal transactive energy schedule $\bm{s}^{*}$.}
     Iteration index $k {\leftarrow} 0$; \\
     Convergence threshold $\epsilon {\leftarrow} 0.000001$; \\
    $\bm{\lambda}^{(0)}_{u,v} {\leftarrow} \bm{0}$, $\bm{p}^{\prime}_{u,v} {\leftarrow} \bm{0}$, $\forall u,v \in \mathcal{U}$; \\
    \While{$\sum_{u \in \mathcal{U}} \sum_{v \in \mathcal{U}} \parallel \bm{p}^{\prime}_{u,v} - \bm{p}_{u,v} \parallel \ge \epsilon$}{
        \For{prosumer $u \in \mathcal{U}$}{
        $\triangleright$ Call smart contract \texttt{Func C} to read $\bm{p}^{\prime}_{u,v}$ and $\bm{\lambda}^{(k)}_{u,v}$;
        
        $\triangleright$ Solves the primal problem numerically;
        
        $\triangleright$ Call smart contract \texttt{Func B} to update $\bm{p}_{u,v}$;
        }
    Smart contract \texttt{Func A} \textbf{do}
    
    $\triangleright$ Wait for all prosumers to update $\bm{p}_{u,v}$;
    
    $\triangleright$ Computes the auxiliary variable $\bm{p}^{\prime}_{u,v}$;
    
    $\triangleright$ Computes the dual variable $\bm{\lambda}_{u,v}$;
    
    $k \leftarrow k+1$;
    }
\end{algorithm}

\end{document}